\documentclass{ws-p8-50x6-00}
\begin{document}

\title{On the origin of Gamma Ray Burst radiation}

\author{Gabriele Ghisellini}

\address{Osserv. Astron. di Brera, Via Bianchi 46, I--23807 Merate,
Italy\\E--mail: gabriele@merate.mi.astro.it}

\maketitle

\abstracts{
In the standard internal shock model, the observed X and gamma--ray 
radiation is assumed to be produced by synchrotron emission.
I will show that there are serious problems with this interpretation, 
calling for other radiation mechanisms, such as quasi--thermal
Comptonization and/or Compton drag processes, or both.
These new ideas can have important consequences on the more general
internal shock scenario, and can be tested by future observations.
}

\section{Introduction}

The most popular interpretation of the origin of the $\gamma$--rays
emitted during the prompt emission of Gamma--Ray Bursts (GRB)
is synchrotron radiation, produced by relativistic leptons accelerated
in shocks formed during the collisions of an inhomogenous 
fireball wind (the so called {\it internal shock scenario},
Rees \& M\'esz\'aros 1992, 1994; Sari \& Piran 1997).

The main evidence for the synchrotron origin of the
prompt radiation is the successful prediction of the
typical energy at which the observed spectrum peaks, in a
$\nu F_\nu$ representation.
However, the very same synchrotron shock scenario inevitably 
predicts very fast radiative cooling of the emitting particles,
with a resulting severe disagreement between the predicted and the observed
spectrum (Ghisellini, Celotti \& Lazzati 2000, see also Cohen et al. 1997;
Sari, Piran \& Narayan 1998; Chiang 1999).

A further general problem of the internal shock scenario is its
low efficiency in transforming bulk kinetic into random energy
and then radiation (Kumar 1999; Lazzati, Ghisellini \& Celotti 2000; 
Spada, Panaitescu \&  M\'esz\'aros 2000).
External shocks (i.e. deceleration of the fireball by the circumburst
matter) should be much more efficient.
Since external shocks are associated with afterglow emission, it
is strange that they are much less energetic than the prompt emission.
It is then compelling to explore alternative possibilities.

\section{A synchrotron origin?}

\subsection{Typical synchrotron frequency}

Assume that two shells with slightly different velocities collide
at some distance $R$ from the explosion site.
In the comoving frame of the faster shell, protons of the other shell
have an energy density 
$U^\prime_{\rm p}=(\Gamma^\prime-1)n^\prime_{\rm p} m_{\rm p}c^2$,
where $\Gamma^\prime\sim 2$ is the bulk Lorentz factor of the slower shell
measured in the rest frame of the other, and $n_{\rm p}$ is the comoving
proton density. 
The magnetic energy density $U^\prime_{\rm B}$ can be amplified to values 
close to equipartition with the proton energy density, 
$U^\prime_{\rm B} = \epsilon_{\rm B} U^\prime_{\rm p}$.
The proton density can be estimated by the kinetic power carried
by the shell :
$L_{\rm s} = 4\pi R^2 \Gamma^2 c n^\prime_{\rm p}  m_{\rm p}c^2$,
yielding $B = (2\epsilon_{\rm B} L_{\rm s}/ c)^{1/2}/( R\Gamma)$.
Also each electron can share a fraction of the available energy, and
if there is one electron for each proton, namely {\it if electron--positron
pairs are not important}, then
$\gamma m_{\rm e} c^2 = \epsilon_{\rm e} m_{\rm p} c^2$.
These simple hypotheses lead to a predicted observed synchrotron frequency 
\begin{equation}
h\nu_{\rm s}\, \sim\, 4\, \epsilon_{\rm e}^2 \epsilon_{\rm B}^{1/2} 
{L_{\rm s, 52}^{1/2} \over R_{13} (1+z) }\,\, {\rm MeV}
\end{equation}
in very good agreement with observations.
Note that the `equipartition coefficients', $\epsilon_{\rm B}$ and 
$\epsilon_{\rm e}$, must be close to unity for the
observed value of $\nu_{\rm peak}$ to be recovered.
In turn this also implies/requires that pairs
cannot significantly contribute to the lepton density.

\subsection{Cooling is fast}

The synchrotron process is a very efficient radiation process.
With the strong magnetic fields required to produce the observed
$\gamma$--rays, the synchrotron cooling time is therefore very short.
As pointed out by Ghisellini, Celotti \& Lazzati (2000), the cooling
time (in the observer frame) can be written as:
\begin{equation}
t_{\rm cool}\,\sim \, 10^{-7} {\epsilon_{\rm e}^3 \Gamma_2 \over 
\nu_{\rm MeV}^2 (1+U_{\rm r}/U_{\rm B})(1+z)}\, \, {\rm s},
\end{equation}
where $U_{\rm r}$ is the radiation energy density.
Note that in Eq. (2) $\epsilon_{\rm B}$ does not appear
(if not for the $U_{\rm r}/U_{\rm B}$ term), and
that $\epsilon_{\rm e}$ is bound to be less than unity.

\subsection{Cooling must be fast}

The smallest variability timescales observed in the
prompt emission of GRBs are of the order of a few milliseconds.
By themselves, these short timescales imply shorter still cooling
timescales: how can the flux decrease if particles do not cool?

\subsection{Predicted synchrotron spectrum}

Since the shortest integration times are of the order of 1 s, the observed 
spectrum is {\it always} the time integrated spectrum produced by
a rapidly cooling particle distribution.

Since $t_{\rm cool}\propto 1/\gamma$, in order to conserve the particle
number, the instantaneous cooling distribution has to satisfy 
$N(\gamma, t) \propto 1/\gamma$.
When integrated over time, the contribution from particles with
different Lorentz factors is `weighted' by their cooling timescale 
$\propto 1/\gamma$.
Therefore the predicted (integrated) flux spectrum is 
\begin{equation}
F_\nu \,  \propto \, 
t_{\rm cool} N(\gamma)\dot\gamma (d\gamma/d \nu) \, \propto \, \nu^{-1/2}
\end{equation}
extending from $\sim$1 keV to $h\nu_{\rm peak}\sim$MeV energies.
We thus conclude that, within the assumptions of the internal
shock synchrotron model, 
a major problem arises in interpreting the observed spectra as 
synchrotron radiation.
We stress that
the ``line of death" of the synchrotron scenario does not correspond
to spectra harder than $F_\nu\propto \nu^{1/3}$, as generally believed,
but to spectra harder than $F_\nu\propto \nu^{-1/2}$, a much more
severe condition: {\it most of the burst spectra do not satisfy it}
(see e.g. Preece et al. 1998; Lloyd \& Petrosian 1999).
Ghisellini, Celotti \& Lazzati (2000) have discussed possible `escape
routes', such as deviations from equipartition, fastly changing magnetic
fields, strong cooling by adiabatic expansion and re--acceleration of
the emitting electrons.
None of these possibilities help. 
The drawn conclusion is that the burst emission is probably produced by 
another radiation process.


\section{Efficiency of internal shocks}

Colliding shells with masses $m_1$, $m_2$,
moving with bulk Lorentz factors
$\Gamma_2>\Gamma_1$ will dissipate part of their
initial bulk kinetic energy with an efficiency $\eta$ given by
\begin{equation}
\eta \,=\, 1-\Gamma_f\, { m_1+m_2\over \Gamma_1 m_1 +\Gamma_2m_2}
\end{equation}
where $\Gamma_f=(1-\beta_f^2)^{-1/2}$ is the bulk Lorentz factor 
after the interaction and is given through 
(see e.g. Lazzati, Ghisellini \& Celotti 1999)
\begin{equation}
\beta_f = {\beta_1\Gamma_1 m_1+ \beta_2\Gamma_2m_2 \over
\Gamma_1m_1+ \Gamma_2 m_2}
\end{equation}
Note that the fraction $\eta$ is not entirely available to produce radiation,
since part of it is in the form of hot protons and magnetic field.
Therefore, as long as the ratio $\Gamma_2/\Gamma_1$ is smaller than a few,
the corresponding efficiency is small.
One could invoke much larger ratios, as in Beloborodov (2000), but in this
case another process, yet to be discussed, becomes relevant, namely the 
Compton drag suffered by a fast shell in the radiation field already 
produced by the collisions of previous shells.
The final outcome is difficult to compute, since it is very likely
that in this situation copious pair production occurs, complicating any
simple description of the process.

\section{Alternatives}

In the previous sections we have discussed two serious problems
for the synchrotron interpretation of the prompt emission and
a more general problem regarding the efficiency of the internal shock
scenario.
We are then motivated to search for alternatives for the main radiation
mechanism of the bursts, and, more generally, to the internal shock scenario.

\subsection{Comptonization}

In the synchrotron shock scenario, one assumes that leptons are accelerated
almost instantaneously to relativistic velocities, and then radiate.
Typical energies are estimated assuming quasi--equipartition
between the different forms of energy (in protons and in magnetic field).
Ghisellini \& Celotti (1999) have investigated instead the alternative
case in which the leptons are accelerated slowly, in a time comparable
to that needed for one shell to cross the other.
The other assumptions of the internal shock scenario (i.e. intermittent
fireball wind, shell mass, typical values of the magnetic field)
where left unchanged with respect to the standard scenario.
What we found is that the accelerated particles in this case
do not attain very large relativistic energies, because the acceleration
and the cooling rates balance at $\gamma<2$.
In these conditions the accelerated particles produce self--absorbed
cyclo--synchrotron emission, and scatter these photons multiple
times to form a quasi--saturated Componization spectrum
(for other models using Comptonization as the main radiation process
see Thompson 1994; Liang 1997, Stern 1999).
Since the typical energy of the particles is small, the resulting
spectrum is very similar to the one produced by a perfect Maxwellian
distribution, even if the actual distribution is different.
In other words, the predicted spectrum is $F_\nu \propto \nu^0$ 
ending in a Wien peak, whose importance with respect to the power law
part of the spectrum is controlled by the Comptonization parameter
$y$ and the particle optical depth $\tau_{\rm T}$.
Since the $\nu^0$ slope is a saturated value, it is appropriate in a large
region of the optical depth -- temperature parameter space.
Furthermore, the temperature $T$ is controlled by electron--positron
pair processes: if $T$ is too large, the produced $\gamma$--rays
interact to form pairs, which share the available energy and lower the
temperature.
This thermostat effect can fix the temperature (at least
in a steady state plasma, so far the only studied in detail)
to less than 50 keV.
The observed spectrum should then peak at $h\nu\sim 3kT\Gamma\sim$
a few MeV.
Since detailed time dependent studies are still lacking,
it is difficult to predict exactly the Comptonized spectrum
and its behavior.
However, for this mechanism to work, it requires an optical depth
of a few and a Comptonization parameter $y\sim$10.

\subsection{Compton drag}

Excesses in the late afterglow optical light curves 
(Bloom et al. 1999; Galama et al. 2000; Bj\"ornsson et al. 2000);
iron lines in emission in X--ray afterglows 
(Piro et al. 1999, 2000, Yoshida et al. 1999; Antonelli et al. 2000) 
and iron absorption edge (Amati et al. 2000)
have revived the interest
in the association between GRBs and Super/Supra/Hyper--Novae.
If the burst explodes soon after or during a supernova explosion,
then the fireball will expand in a dense photon environment,
and it can be decelerated by the Compton drag effect
(Lazzati et al., 2000).
In this case there will be $direct$ transformation of bulk energy
into radiation, without the need to randomly accelerate the
emitting particles.
Furthermore, the fireball starts to produce radiation even when 
optically thick.

The required existing seed radiation field is indeed very large, 
such as the one produced by the walls of the funnel of an hypernova
or by the just born remnant of a supernova exploded hours before
the bursts.
It is instead more problematic to decelerate the fireball if the 
seed photon field is produced by the remnant exploded a few 
months before the burst, because in this case the characteristic 
size of the remnant is already large, and the radiation is diluted.

\begin{figure}[t]
\epsfxsize=33pc 
\epsfbox{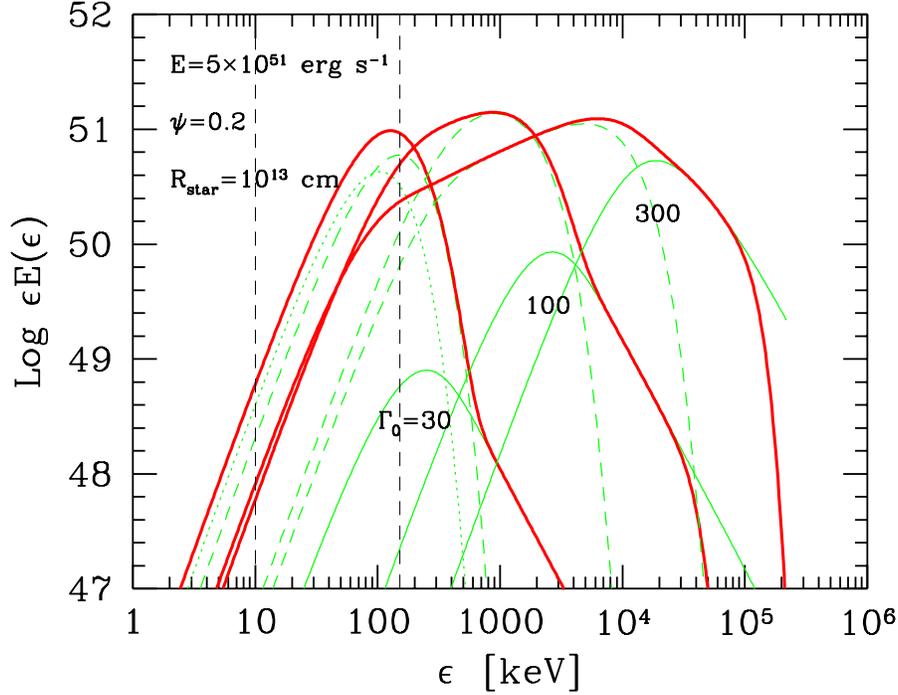} 
\vskip -4 true cm
\caption{The spectra predicted by a fireball dragged in the dense
photon environment of an hypernova (From Ghisellini et al., 2000).}
\end{figure}

The characteristic frequency of the scattered radiation is 
$\nu\sim 2\Gamma^2 \nu_0$ where $\nu_0$ is the typical frequency
of the incident seed photon.
Most of the emission is produced when the fireball has an optical depth
around unity, because this corresponds to the largest volume within 
which most of the ambient photons are scattered.
If the process is efficient it can decelerate the fireball, which
then scatter radiation with different $\Gamma$ factors.
Furthermore, in the case of a GRB born in a funnel of an hypernova,
the walls of the funnel will be characterized by a distribution
of temperatures (hotter in the inner parts).
Therefore, even if the locally produced spectrum is blackbody--like,
the time--integrated radiation is a superposition of blackbody 
spectra, each one characterized by a different effective observed
temperature $T_{\rm obs}\sim 2T\Gamma^2$.
Some examples of computed spectra in these conditions are plotted
in Fig. 1, for a fireball of total energy $5\times 10^{51}$ erg,
expanding in a funnel of semiaperture angle $\psi=0.2$, in a star
of radius $R_{\rm star}=10^{13}$ cm.

The initial bulk Lorentz factor is $\Gamma$=30, 100, 300 in the
three examples shown.
Since the fireball energy is kept fixed, a small $\Gamma$ correspond
to a large fireball mass and viceversa.
The efficiency of the process is then dependent of $\Gamma$ since
the energy loss rate is $\propto \Gamma^2$.

For the adopted parameters, the spectrum peaks at 0.1--10 MeV,
has a steep high energy tail, and a low energy $\nu^2$ slope.
While this is in qualitative agreement with what observed,
we point out that the low energy slope of this spectrum (the $\nu^2$
part) is a clear prediction of this model.
It is in fact produced quite naturally, being the 
blackbody input spectrum boosted by the square of the Lorentz
factor that the fireball has soon after it has become transparent.
Instead, other models would have difficulties in explaining it,
since synchrotron self--absorption occurs at much lower frequencies.
In Fig. 1 we have indicated the energy range between 10 and 150 keV,
appropriate for the high energy instrument onboard the SWIFT satellite,
which can be particularly revealing in this respect.

\subsection{Both?}

Consider a fireball consisting of several shells (as an approximation
of an inhomogeneous wind) crossing the dense photon environment of
an hypernova funnel.
If the process is efficient, the first shells will decelerate 
as a result of the intense Compton drag, but doing so, they will free
the funnel cavity of photons.
Therefore the later shells will suffer less Compton drag, 
especially along the axis of the funnel.
This have important consequences:
\begin{itemize}
\item Internal shocks between first and later shells will develop
{\it even if the initial velocities are equal}.

\item Later shells will be faster along the axis than close to the funnel
walls, where they can be dragged by the newly produced seed photons.

\item Instabilities can occur in the funnel, more likely close to 
funnel walls, leading to strong shocks with dense material.
This in turn favors inverse Compton as main radiation process.
\end{itemize}

All this is admittedly very qualitative, but worth to be investigated.

\section{Conclusions}

The synchrotron interpretation of the prompt emission of GRBs
faces a severe problem: the synchrotron process is in fact
very efficient, implying very short cooling timescales,
shorter than the dynamical time (the light crossing
time across the shell width).
All the spectra obtained so far are time integrated over millions 
of cooling times, and are then produced by a cooling particle 
population, whose expected spectrum is much softer than observed.
On the contrary the afterglow emission can well be due to the 
synchrotron process: in this case the much lower value of the 
typical magnetic field yields longer cooling timescales with 
incomplete cooling of the particle distribution.

We have explored alternatives such as Comptonization by a quasi--thermal 
particle distribution, whose temperature is kept small by 
electron--positron feedbacks, and the Compton drag process,
requiring a very dense pre--existing photon field and the association
of GRBs with just exploded (or about to explode) supernovae.

\section*{Acknowledgments}
Davide Lazzati and Annalisa Celotti are gratefully thanked for many discussions
and years of fruitful collaboration.

\end{document}